\pdfoutput=1 
\documentclass[11pt,a4paper]{article}
\usepackage[utf8]{inputenc}

\usepackage{jheppub}
\usepackage[table]{xcolor}
\usepackage{colortbl}
\RequirePackage{ifpdf} 
\usepackage{amsmath} 
\usepackage{mathtools}
\usepackage{cases}
\usepackage{comment}
\usepackage[colorinlistoftodos]{todonotes}
\usepackage{adjustbox}
 
\usepackage{pstricks}
\usepackage[final]{pdfpages} 
\usepackage{ifpdf} 
\usepackage{slashed}
\usepackage{color}
\usepackage{xcolor}
\definecolor{urlblue}{rgb}{0.2,0.4,0.7}
\definecolor{citegreen}{rgb}{0,0.6,0.2}
\definecolor{linkred}{rgb}{0.9,0.2,0.1}
\usepackage{float}
\usepackage[caption = false]{subfig}
\usepackage{graphics}
\usepackage{etoolbox} % Load before 'breqn' package to make 'lineno' work
\usepackage{fixmath}
\usepackage{psfrag}
\graphicspath{{Figure/}}

\usepackage{notoccite} % If you have \cite commands in \section-like commands, or in \caption, the
\usepackage{amsfonts}
\usepackage{autobreak}
\usepackage{marginnote}

\usepackage{nccmath}
    {\end{pmatrix}\end{medsize}}%

\usepackage{tikz}
\usetikzlibrary{positioning,arrows}
\usetikzlibrary{decorations.pathmorphing}
\usetikzlibrary{decorations.markings}
\usetikzlibrary{shapes.geometric}
\tikzset{
    vector/.style={decorate, decoration={snake}, draw},
    provector/.style={decorate, decoration={snake,amplitude=2.5pt}, draw},
    antivector/.style={decorate, decoration={snake,amplitude=-2.5pt}, draw},
    fermion/.style={draw=black, postaction={decorate},decoration={markings,mark=at position .55 with {\arrow[draw=black]{>}}}},
    fermionbar/.style={draw=black, postaction={decorate},
                       decoration={markings,mark=at position .55 with {\arrow[draw=black]{<}}}},
    fermionnoarrow/.style={draw=black},
    gluon/.style={decorate, draw=black,decoration={coil,amplitude=4pt, segment length=5pt}},
    scalar/.style={dashed,draw=black, postaction={decorate},decoration={markings,mark=at position .55 with {\arrow[draw=black]{>}}}},
    scalarbar/.style={dashed,draw=black, postaction={decorate},decoration={markings,mark=at position .55 with {\arrow[draw=black]{<}}}},
    scalarnoarrow/.style={dashed,draw=black},
    electron/.style={draw=black, postaction={decorate},decoration={markings,mark=at position .55 with {\arrow[draw=black]{>}}}},
    bigvector/.style={decorate, decoration={snake,amplitude=4pt}, draw},
}

\title{Two-loop non-planar four-point topology with massive internal loop} 

\author{Taushif Ahmed$^{a}$, Ekta Chaubey$^{b}$, Mandeep Kaur$^{c}$ and Sara Maggio$^{b}$}
\emailAdd{taushif.ahmed@ur.de, eekta@uni-bonn.de, mandeepkaur.iiser@gmail.com, smaggio@uni-bonn.de}
\affiliation{
$^a$Institute for Theoretical Physics, University of Regensburg, 93040 Regensburg, Germany 
\\$^b$Bethe Center for Theoretical Physics, Universitaet Bonn, 53115 Bonn, Germany
\\$^c$Indian Institute of Science Education and Research Mohali,\\
Knowledge City, Sector 81, SAS Nagar, Manauli, Punjab 140306, India}

\preprint{}

\abstract{
 We study a set of two-loop non-planar master integrals needed for the NNLO QCD corrections to diphoton and dijet production at hadron colliders. The top-sector topology contains an internal massive fermion loop and is known to contain elliptic curves. Leveraging the method of differential equations, we provide a comprehensive discussion for deriving an $\epsilon$-factorized differential equation related to the most intricate sector within the Feynman integral family. Despite the dependence on multiple scales and the presence of two elliptic sectors, we demonstrate how to leverage the properties of their maximal cuts and the factorization of the Picard-Fuchs operator to deal with the complexity of the analytic computation. In particular, we construct a transformation matrix that brings the differential equations into a format enabling the convenient expression of analytic results in terms of Chen's iterated integrals.
 }

\begin{document}
\allowdisplaybreaks[4]
\unitlength1cm
\maketitle
\flushbottom

%************
% Definition
%************

\def\D{{\cal D}}
\def\DD{\overline{\cal D}}
\def\g{\overline{\cal C}}
\def\gm{\gamma}
\def\M{{\cal M}}
\def\ep{\epsilon}
\def\epm1{\frac{1}{\epsilon}}
\def\epm2{\frac{1}{\epsilon^{2}}}
\def\epm3{\frac{1}{\epsilon^{3}}}
\def\epm4{\frac{1}{\epsilon^{4}}}
\def\unM{\hat{\cal M}}
\def\ashat{\hat{a}_{s}}
\def\asmur{a_{s}^{2}(\mu_{R}^{2})}
\def\sigbar{{{\overline {\sigma}}}\left(a_{s}(\mu_{R}^{2}), L\left(\mu_{R}^{2}, m_{H}^{2}\right)\right)}
\def\sigbarn{{{{\overline \sigma}}_{n}\left(a_{s}(\mu_{R}^{2}) L\left(\mu_{R}^{2}, m_{H}^{2}\right)\right)}}
\def\unas{ \left( \frac{\hat{a}_s}{\mu_0^{\epsilon}} S_{\epsilon} \right) }
\def\rnM{{\cal M}}
\def\bt{\beta}
\def\cD{{\cal D}}
\def\cC{{\cal C}}
\def\ca{\text{\tiny C}_\text{\tiny A}}
\def\cf{\text{\tiny C}_\text{\tiny F}}
\def\ct{{\red []}}
\def\sv{\text{SV}}
\def\murOmu{\left( \frac{\mu_{R}^{2}}{\mu^{2}} \right)}
\def\bb{b{\bar{b}}}
\def\bt0{\beta_{0}}
\def\bt1{\beta_{1}}
\def\bt2{\beta_{2}}
\def\bt3{\beta_{3}}
\def\gm0{\gamma_{0}}
\def\gm1{\gamma_{1}}
\def\gm2{\gamma_{2}}
\def\gm3{\gamma_{3}}
\def\nn{\nonumber}
\def\l{\left}
\def\r{\right}
\def\T{{\cal Z}}    
\def\U{{\cal Y}}

\def\nn{\nonumber\\}
\def\ep{\epsilon}
\def\T{\mathcal{T}}
\def\V{\mathcal{V}}

\def\qgraf{{\fontfamily{qcr}\selectfont
QGRAF}}
\def\python{{\fontfamily{qcr}\selectfont
PYTHON}}
\def\form{{\fontfamily{qcr}\selectfont
FORM}}
\def\reduze{{\fontfamily{qcr}\selectfont
REDUZE2}}
\def\kira{{\fontfamily{qcr}\selectfont
Kira}}
\def\litered{{\fontfamily{qcr}\selectfont
LiteRed}}
\def\fire{{\fontfamily{qcr}\selectfont
FIRE5}}
\def\air{{\fontfamily{qcr}\selectfont
AIR}}
\def\mint{{\fontfamily{qcr}\selectfont
Mint}}
\def\hepforge{{\fontfamily{qcr}\selectfont
HepForge}}
\def\arXiv{{\fontfamily{qcr}\selectfont
arXiv}}
\def\Python{{\fontfamily{qcr}\selectfont
Python}}
\def\ginac{{\fontfamily{qcr}\selectfont
GiNaC}}
\def\polylogtools{{\fontfamily{qcr}\selectfont
PolyLogTools}}
\def\anci{{\fontfamily{qcr}\selectfont
Finite\_ppbk.m}}

\newcommand{\dis}{}
\newcommand{\overbar}[1]{mkern-1.5mu\overline{\mkern-1.5mu#1\mkern-1.5mu}\mkern
1.5mu}

%%%%%%%%%%%%%%%%%%%%%%%%%%%%%%%%%%%%%%
\begin{comment}

{\todo[inline]{I think in the title we should mention two-loop non-planar 4-point master integrals or master integrals for (process name) or something for more clarity for e.g. Master integrals for two-loop non-planar QCD corrections to diphoton and dijet production with full top-quark mass dependence }}
    
\end{comment}
\section{Introduction}
\label{sec:intro}
% importance of MI
Over the past few decades, an enormous amount of progress has been made in under-
standing Feynman integrals and their corresponding functional spaces. This progress is
necessitated by the growing need for precise theoretical calculations to effectively confront
the experimental data collected at the present and potential future colliders. With the upgraded Large Hadron Collider (LHC) operating at higher luminosity and the promise of even higher energy reach for future colliders, the contributions of massive quark loops to scattering processes have become increasingly significant.
The multi-loop amplitudes governing these scattering processes can generally be expressed in terms of an independent set of integrals called master integrals (MIs). In the realm of perturbative Quantum Field Theory (pQFT), obtaining full analytical results for processes involving heavy particles as mediators have been limited to initial orders due to the non-availability of the results for involved complicated integrals. In this article, we address such a scenario by focusing on the analytical computation of a specific class of two-loop Feynman integrals crucial for higher-order corrections to two very important processes at the LHC. Specifically, we present the analytic computation of the two-loop Feynman integrals associated with a multi-scale non-planar topology, which appears in NNLO QCD corrections to the diphoton and dijet production processes.  

Owning to the relevance of diphoton and dijet production processes in searches for physics beyond the Standard Model (SM){~\cite{ATLAS:2015nsi,CMS:2016gsl,CMS:2018dqv}} and in testing the SM predictions, obtaining analytic expressions for their production cross-sections is crucial. These expressions are essential for improving prediction accuracy and enabling more accurate comparisons with the experimental data. However, the presence of non-trivial massive Feynman integrals with dependence on multiple kinematic scales pose significant challenges in obtaining the analytic expressions for their production cross-sections at the two-loop level.  
 The state-of-the-art studies of analytic computation of two-loop integrals contributing to NNLO QCD corrections to the dijet and diphoton production processes are as follows. The analytic results for the two-loop massive planar integrals have been known for quite some time{~\cite{Caron-Huot:2014lda,Becchetti:2017abb,H:2023wfg, Becchetti:2023wev}}. These results are obtained in terms of Chen's iterated integrals and, whenever possible, expressed in terms of Goncharov's polylogarithms (GPLs). The integrals belonging to a non-planar subtopology keeping the full dependence on the internal top-quark mass have been computed in refs.~\cite{vonManteuffel:2017hms} and are identified to be involving an elliptic curve. Other subtopologies in the context of Higgs boson have been calculated in refs.~\cite{Aglietti:2006tp,Anastasiou:2006hc}. The lack of analytic results for one of the non-planar topologies, as shown in figure.~\ref{fig:familyF}, prevents us from getting the full two-loop amplitude for dijet and diphoton production in a closed analytic form. In refs.~\cite{Xu:2018eos,Wang:2020nnr}, the integrals belonging to this topology were shown to be elliptic and are evaluated numerically. Recently, the results for all non-planar integrals contributing to the two-loop form factors for diphoton production have been computed using semi-analytical methods~\cite{Hidding:2020ytt, Becchetti:2023wev}. Despite all these efforts, full analytic results for the non-planar elliptic integral family are still missing from the literature.

In this article, we study all the two-loop elliptic sector integrals belonging to the non-planar topology shown in figure.~\ref{fig:familyF}, keeping the full dependence on the top-quark mass ($m_t$) running in the loop with the aim of obtaining their analytic results. In the first step, we perform the integration-by-parts (IBP)~\cite{TKACHOV198165,CHETYRKIN1981159} reduction for the given topology to get the minimal set of scalar integrals called master integrals (MIs) using {\tt{KIRA}}~\citep{Maierhofer:2017gsa} and {\tt{FIRE}}~\citep{Smirnov:2014hma}. Then, we set up the system of differential equations for these master integrals~\citep{Kotikov:1990kg,Kotikov:2003tc,Remiddi:1997ny,GEHRMANN2000485,Argeri:2007up} and aim to bring it into an $\epsilon$-factorized form~\citep{Henn:2013pwa,Henn:2014qga} by suitable basis choice, where the dimensional regulator $\epsilon$ is related to space-time dimension through $d=4-2\epsilon$. Since the subsector integrals are well-documented in the literature, obtaining an $\epsilon$-factorized system of differential equations for them is straightforward. Additionally, these sub-sector integrals can be expressed either in terms of multiple polylogarithms (MPLs)~\cite{Goncharov:1995tdt,Goncharov:2001iea,Remiddi:1999ew} or elliptic multiple polylogarithms~\cite{Broedel:2019hyg}. However, for the top-sector integrals, characterized by the presence of an elliptic curve, the results cannot be expressed in terms of MPLs. The considered topology is particularly intricate because it not only depends on multiple kinematic scales but also contains an elliptic subtopology. Despite the current prominence of such integrals in research, there are no universal algorithms applicable to every case, and each new topology requires specific attention ~\citep{Adams:2013nia,Adams:2014vja,Adams:2018kez,Duhr:2021fhk,Laporta:2004rb,Muller-Stach:2012tgz,Bloch:2013tra,Adams:2015gva,Adams:2015ydq,Adams:2016xah,Adams:2017tga,Bogner:2017vim,Broedel:2017kkb,Adams:2018yfj,Broedel:2018iwv,Adams:2018bsn,Broedel:2018qkq,Honemann:2018mrb,Bogner:2019lfa,Broedel:2019hyg,Broedel:2019kmn,Weinzierl:2019pfw,Adams:2017ejb,Marzucca:2023gto,Giroux:2024yxu,Gorges:2023zgv, Delto:2023kqv}. 
To cast the system of differential equations into the $\epsilon$-form for the top-sector involving an elliptic curve, we have generalised the technique discussed in~\citep{Pogel:2022vat} for the basis transformation.
Finally, we discuss how to use our choice of basis to express the analytic results of the top-sector master integrals at each order in $\epsilon$ through Chen's iterated integrals~\citep{Chen:1977oja} over elliptic one-forms.

The paper is organized as follows: In section~\ref{sec:set-up}, we introduce our notation for the kinematics and integral family, as well as discuss our choice of master integral basis for the topology under consideration. Section~\ref{sec:Elliptic-top-sector} is devoted to a detailed discussion on establishing the $\epsilon$-form of the differential equation system for the most intricate elliptic top-sector, utilizing the maximal cut and factorization of Picard-Fuchs operator information. Finally, we discuss how to express all the integrals belonging to this Feynman integral family through Chen's iterated integrals.

\section{The setup}
\label{sec:set-up}
\begin{figure}[htbp]
\begin{center}
\includegraphics[width = 2.1in, height=1.3in]{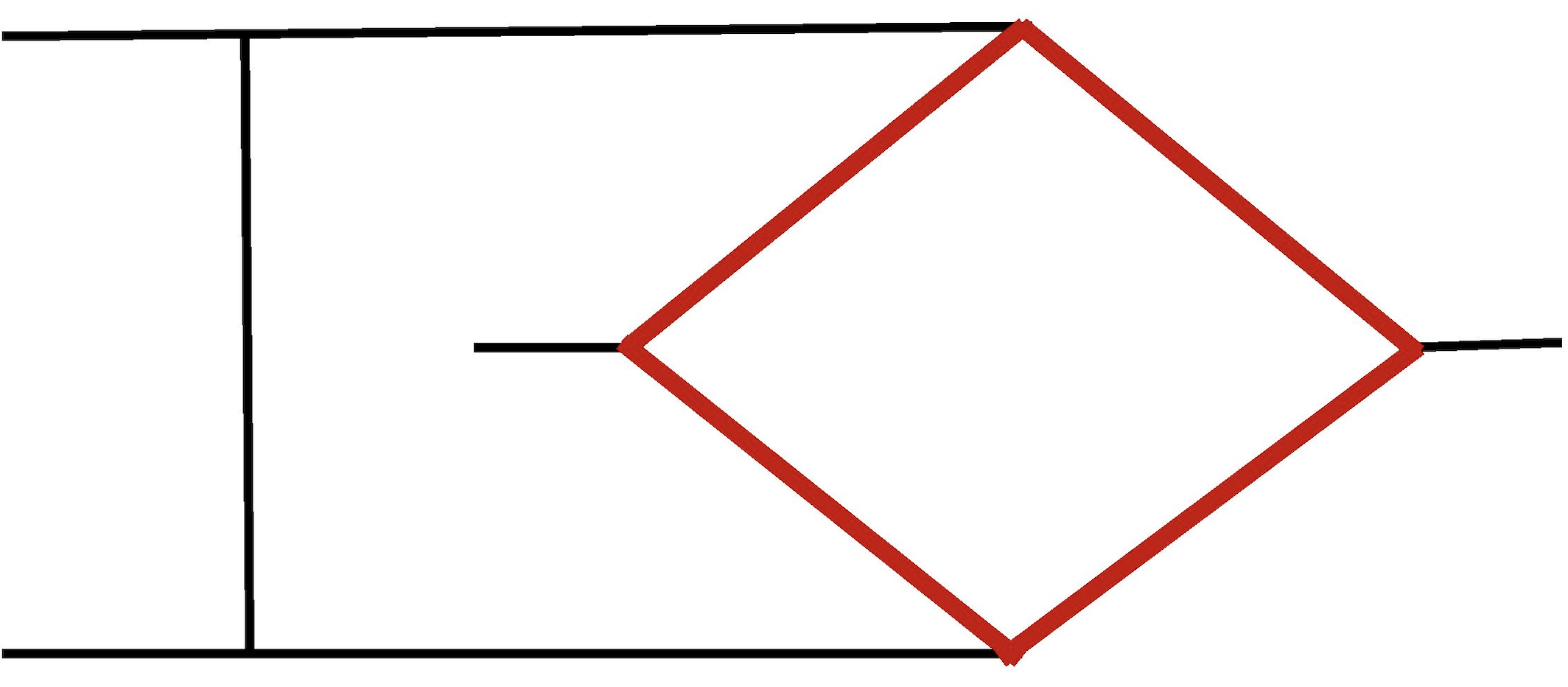}
\caption{The non-planar topology where the red and black lines respectively denote massive and massless propagators.}
\label{fig:familyF}
\end{center}
\end{figure}

We consider a generic four-point scattering process involving massless external particles with momenta $p_1,\cdots,p_4$, all of them we regard as incoming. The external momenta obey the following kinematics: $\sum_{i=1}^4 p_i = 0,\; p_i^2=0\;\text{for}\;i=1,2,3,4$. We define the following abbreviations to represent the independent set of Mandelstam variables
\begin{align}
\label{eq:Mandelstam}
s = (p_1+p_2)^2, \qquad  t=  (p_1+p_3)^2.
\end{align}
Let us define a set of two-loop non-planar Feynman integrals through
\begin{align}
\label{eq:define-int}
I_{a_1,\cdots,a_9} = \left( \frac{e^{\epsilon \gamma_E}}{i\pi^{\frac{d}{2}}} \right)^2 \int \prod_{i=1}^2 d^dk_i \frac{D_8^{a_8} D_9^{a_9}}{D_1^{a_1} D_2^{a_2} D_3^{a_3}D_4^{a_4}D_5^{a_5}D_6^{a_6}D_7^{a_7}}\,, \quad a_j \in \mathbb{Z}\,.
\end{align}
We represent the loop momenta through $k_i$, Euler-Mascheroni constant by $\gamma_E$ and the inverse propagators by $D_j=q_j^2-m_j^2+i 0^+$ with the parameters $q_j$ and $m_j$ respectively denoting the momentum and mass. In this article, we focus on a set of Feynman integrals associated with the topology shown in figure~\ref{fig:familyF}.
The corresponding integral family is constructed out of the following set of inverse propagators
\begin{align}
    \big\{ &k_1^2, (k_1-p_1)^2, (k_1-p_1-p_2)^2, k_2^2-m_t^2, (k_2-p_1-p_2-p_3)^2-m_t^2,\nonumber\\
    &(k_1-k_2)^2-m_t^2, (k_1-k_2+p_3)^2-m_t^2, (k_1-p_1-p_2-p_3)^2, (k_2+p_1)^2 -m_t^2\big\}\,,
\end{align}
where $m_t$ denotes the mass of top quark appearing in the loop.

\subsection{Choice of master integrals}
\label{sec:master-integrals}

By using the integration-by-parts~\cite{TKACHOV198165,CHETYRKIN1981159} identities for the given topology with the help of {\tt{KIRA}}~\citep{Maierhofer:2017gsa} and {\tt{FIRE}}~\citep{Smirnov:2014hma}, we obtain a basis of 36 master integrals denoted by $\vec{I}$. We then set up a system of differential equations for these master integrals with respect to the Mandelstam variables defined in equation~(\ref{eq:Mandelstam}). Utilizing IBP relations, the differential equations take the form:
\begin{equation}
    d\Vec{I}=A\Vec{I},
    \label{eq:ibpvector}
\end{equation}
where $A$ is a $36 \times 36$ matrix-valued one-form depending on the Mandelstam variables and the space-time dimensions ($d= 4-2\epsilon$). 

Since the basis choice for the master integrals is not unique, we can perform a basis transformation
\begin{equation}
    \Vec{{J}}=U\Vec{I},
\end{equation}
which leads to a transformed system of differential equations
\begin{equation}
    d\Vec{{J}}=\Tilde{A}\Vec{{J}},
     \label{DE}
\end{equation}
where,
\begin{equation}
    \Tilde{A}=UA\;U^{-1}-UdU^{-1}.
    \label{trafoDE}
\end{equation}
If we can find a transformation matrix $U$ such that the transformed differential equation attains the form
\begin{equation}
    d\Vec{{J}}=\epsilon \Tilde{A}\Vec{{J}},
    \label{epsDE}
\end{equation}
such that the matrix-valued one-form $\Tilde{A}$ does not have any $\epsilon$-dependence, then the analytic result of $\Vec{{J}}$ can be systematically expressed in terms of Chen's iterated integrals order by order in $\epsilon$. Additionally, each entry at a given $\epsilon$ order will have a uniform transcendental weight. We may slightly relax this condition of complete $\epsilon$-factorization and have the form of $\Tilde{A}$ as 
\begin{equation}
    \Tilde{A}=\Tilde{A}^{(0)}+\epsilon \Tilde{A}^{(1)},
    \label{linearDE}
\end{equation}
where $\Tilde{A}^{(0)}$ is strictly lower triangular and both $\Tilde{A}^{(0)}$ and $\Tilde{A}^{(1)}$ are independent of $\epsilon$~\cite{Adams:2018kez}. 
This form of the transformed matrix $\tilde{A}$ maintains the property that we can still express the results for $\vec{J}$ in terms of iterated integrals at every order in $\epsilon$, although there might be a mixing of functions with different \textit{length} at a particular order in $\epsilon$. Here length is defined as the number of integrations that needs to be performed iteratively. 
With the modern technology of using differential equations for computing Feynman integrals, often the bottleneck is the construction of a transformation matrix that brings our initial ``not so good set of differential equations'' to a better form, such as the canonical form in equation~(\ref{epsDE}) or the special linear form in equation~(\ref{linearDE}). In this work, we highlight how to obtain both these forms for the top-sector.

Instead of working with basis $\vec{I}$ in equation~(\ref{eq:ibpvector}), guided by the works in~\citep{Caron-Huot:2014lda,Becchetti:2017abb,H:2023wfg,vonManteuffel:2017hms}, we choose the following definitions of the master integrals to set up our differential equations, where we have used $m_t^2=1$ for convenience: 
\begin{align}
\label{eq:startingDE}
J_{1}\;&= \; \epsilon^2 \;\; \textbf{D}^{-} I_{000010100},\nonumber\\
J_{2}\;&= \; \epsilon^2 \;\; \textbf{D}^{-} I_{101000100},\nonumber\\
J_{3}\;&= \; \epsilon^2\; \sqrt{(-4+s)\; s}\left[ I_{002201000}+\frac{1}{2} I_{001202000}\right],\nonumber\\
J_{4}\;&= \; \epsilon^2\; \sqrt{s + t} \sqrt{4 + s + t}\left[ I_{020021000}+\frac{1}{2} I_{010022000}\right],\nonumber\\
J_{5}\;&= \; \epsilon^2\;  \sqrt{(4 - t)\;t}\left[ I_{020200100}+\frac{1}{2} I_{010200200}\right],\nonumber\\
J_{6}\;&= \; \frac{1}{2} \epsilon^2 s \left[I_{001202000} + 
   2 (I_{002201000} + 
      \textbf{D}^{-} I_{001101100})\right],\nonumber\\
J_{7}\;&= \; \epsilon^3\; s\; I_{002101100},\nonumber\\
J_{8}\;&= \; \epsilon^2 \;\; (s + t)\; \left[\frac{1}{2}I_{010022000}+I_{020021000} + \textbf{D}^{-} I_{010011100}\right],\nonumber\\
J_{9}\;&= \; \epsilon^3\; (s + t)\;  I_{020011100},\nonumber\\
J_{10}\;&= \; \frac{1}{2}\; \epsilon^2  \; t\;\left[(I_{010200200} + 
   2 (I_{020200100}+\; \textbf{D}^{-} I_{010101100})\right],\nonumber\\
J_{11}\;&= \; \epsilon^3\; t \; I_{020101100},\nonumber\\
J_{12}\;&= \; \epsilon^4\; s \; I_{001111100},\nonumber\\
J_{13}\;&= \; \epsilon^4\; s \; I_{101011100},\nonumber\\
J_{14}\;&= \; \epsilon^3\; s \; I_{101012000},\nonumber\\
J_{15}\;&= \; \epsilon^2\; \sqrt{-s(4 + s)}\; \left[ -s\;I_{201012000} + \epsilon \frac{\textbf{D}^{-} I_{000010100}}{2+4\epsilon}\right],\nonumber\\
J_{16}\;&= \; \epsilon^4\; (s + t)\; I_{011101100},\nonumber\\
J_{17}\;&=  -\epsilon^3\; \sqrt{s \;t \;(-4\;s + s\;t - 4\;t)} \; I_{011201100},\nonumber\\
J_{18}\;&= \; \epsilon^4\; t\; I_{011111000},\nonumber\\
J_{19}\;&= -\epsilon^3\; \sqrt{s\; (s + t)\; (s^2  -4\;t + s\; t)} \;I_{011112000},\nonumber\\
J_{20}\;&= \; \epsilon^4\; s\; I_{010111100},\nonumber\\
J_{21}\;&= \; \epsilon^3\; \sqrt{t\; (s + t)\; ( t^2-4\;s + s\;t )}\; I_{020111100},\nonumber\\
J_{22}\;&= \; \epsilon^3\; s\; (I_{010121100} + I_{010211100}),\nonumber\\
J_{23}\;&= \; \epsilon^4\; s\; (s + t)\; I_{111011100},\nonumber\\
J_{24}\;&= \; \epsilon^4\; s\;t\;  I_{111101100},\nonumber\\
J_{25}\;&= \; \epsilon^3\;s \;\sqrt{(-4 + t)\; t} \;I_{111100200},\nonumber\\
J_{26}\;&= \; \epsilon^3\;s \;((-1 + 2 \epsilon)\; I_{111100100} - (-4 + t)\;I_{111100200}),\nonumber\\
J_{27}\;&= \; \epsilon^4\;  \sqrt{s}\; \sqrt{t}\; \sqrt{s + t} \;I_{011111100},\nonumber\\
J_{28}\;&= \; \frac{1}{4}\;\epsilon^3\;\left[s\; (s + t)\; (I_{011112000}+ 4 I_{011112100}) + 
  s\; t \;I_{011201100} - 
  t\; (s + t)\; I_{020111100}\right],\nonumber\\
J_{29}\;&= \; \epsilon^3\;s \;\sqrt{s + t}\; \sqrt{4 + s + t} \;I_{111012000},\nonumber\\
J_{30}\;&= \; \epsilon^3\;s \left[(-1 + 2 \epsilon) \;I_{111011000} + (4 + s + t)\; I_{111012000}\right],\nonumber\\
J_{31}\;&= -\frac{\pi\; s^2 \; \epsilon^4 }{2 \;\text{K}\big(\frac{-16}{s}\big)} I_{101111100},\nonumber\\
J_{32}\;&=\frac{\epsilon^3 s}{\pi} \bigg[(1 + 
      2 \epsilon)\; \bigg(s\; \text{E}\big(-\frac{16}{s}\big) - (16 + s)\; \text{K}\big(-\frac{16}{s}\big)\bigg) I_{1, 0, 1, 1, 1, 1, 1, 0, 0} \nonumber\\&\;- 
   8\; (16 + s)\; \text{K}\big(-\frac{16}{s}\big) I_{1, 0, 1, 2, 1, 1, 1, 0, 0}\bigg]\nonumber\\
J_{33}\;&= \; \epsilon^4\;  I_{111111100},\nonumber\\
J_{34}\;&= \; \epsilon^4\;  I_{11111110-2},\nonumber\\
J_{35}\;&= \; \epsilon^4\;  I_{11111110-1},\nonumber\\
J_{36}\;&= \; \epsilon^4\;  I_{1111111-1-1}\,.
\end{align}
Here $\textbf{D}^{-}$ denotes the dimension shift operator which shifts the space-time dimension $d$ by two as
\begin{align}
   \textbf{D}^{-} I_{a_1,\cdots,a_9} (d)= I_{a_1,\cdots,a_9} (d-2),
\end{align}
whereas $K(\cdot)$ and $E(\cdot)$ are respectively the complete elliptic integral of the first and second kind,
\begin{equation}
    K(k^2)=\int_0^1\frac{dt}{\sqrt{(1-t^2)(1-k^2t^2)}}, \qquad E(k^2)=\int_0^1 \sqrt{\frac{1-k^2t^2}{1-t^2}} dt.
\end{equation}
With this basis choice, the differential equations readily assume a canonical form, as in equation~(\ref{epsDE}) for the first 31 integrals, and for $J_{32}$ it takes on a special linear form, as in equation~(\ref{linearDE}), that can be effortlessly transformed into the canonical one through a simple transformation involving the primitive, as outlined in~\cite{Adams:2017tga}. Moreover, the integrals $J_{33}$-$J_{36}$ in the top sector, featuring all seven propagators, along with $J_{31}$-$J_{32}$, constitute elliptic sectors. In the following sections, we explain the main ingredients of our work, tackling the analytic complexity of the top sector due to the elliptic nature of the Feynman integral. 

\section{The elliptic top-sector}
\label{sec:Elliptic-top-sector}
As previously mentioned, all sub-sector integrals associated with the non-planar family~\ref{fig:familyF} have been known analytically for quite some time. Nevertheless, writing the analytic results for the full non-planar topology is challenging due to the presence of elliptic curves, which extend beyond the well-understood class of polylogarithmic functions. The presence of multiple kinematic scales adds further to the analytic complexity. This makes the analytic computation of the entire integral family considerably more challenging compared to many other multi-scale two-loop Feynman integrals.
Therefore, in the following sections, we focus primarily on the top sector, i.e. integrals $J_{33}$-$J_{36}$ and provide a highly detailed explanation of the steps involved in transforming the corresponding differential equations in a way that facilitates the straightforward articulation of analytic results. The ideas presented here can be easily generalized to other multi-variate elliptic Feynman integrals involving multiple elliptic sectors. 

\subsection{The study of maximal cuts}
Maximal cuts are one of the primary tools that can be utilized to analyse the algebraic complexity of multi-loop, multi-scale Feynman integrals before solving them explicitly~\cite{BAIKOV1997347,Primo:2017ipr,Frellesvig:2017aai}. In this context, `cutting' a propagator means forcing the propagating particle to be on-shell. This can be achieved for the propagators raised to power one by simply replacing each of them with a Dirac $\delta$-function. A maximal cut of a diagram corresponds to the simultaneous cutting of all its propagators. Mathematically, a maximal cut corresponds to taking the $n$-fold residue of the corresponding integrand in the complex plane, where $n$ is the number of propagators. Furthermore, it is known that the maximal cuts of any given set of master integrals are the solutions for the corresponding homogeneous system of differential equations~\cite{Primo:2016ebd,Adams:2018kez}. Therefore, the maximal cut information of a given Feynman integral is useful to obtain the integral representation for the homogeneous solution of the differential equation
satisfied by that integral. The computation of the maximal cut can be easily carried out in the so-called Baikov representation~\cite{Baikov:1996iu} using the loop-by-loop approach \cite{Frellesvig:2017aai}.

\begin{figure}[htbp]
\begin{center}
%\vspace{-1cm}
\subfloat[(a)]{\adjustbox{valign=c}{\includegraphics[width = 2.5in, height=1.25in]{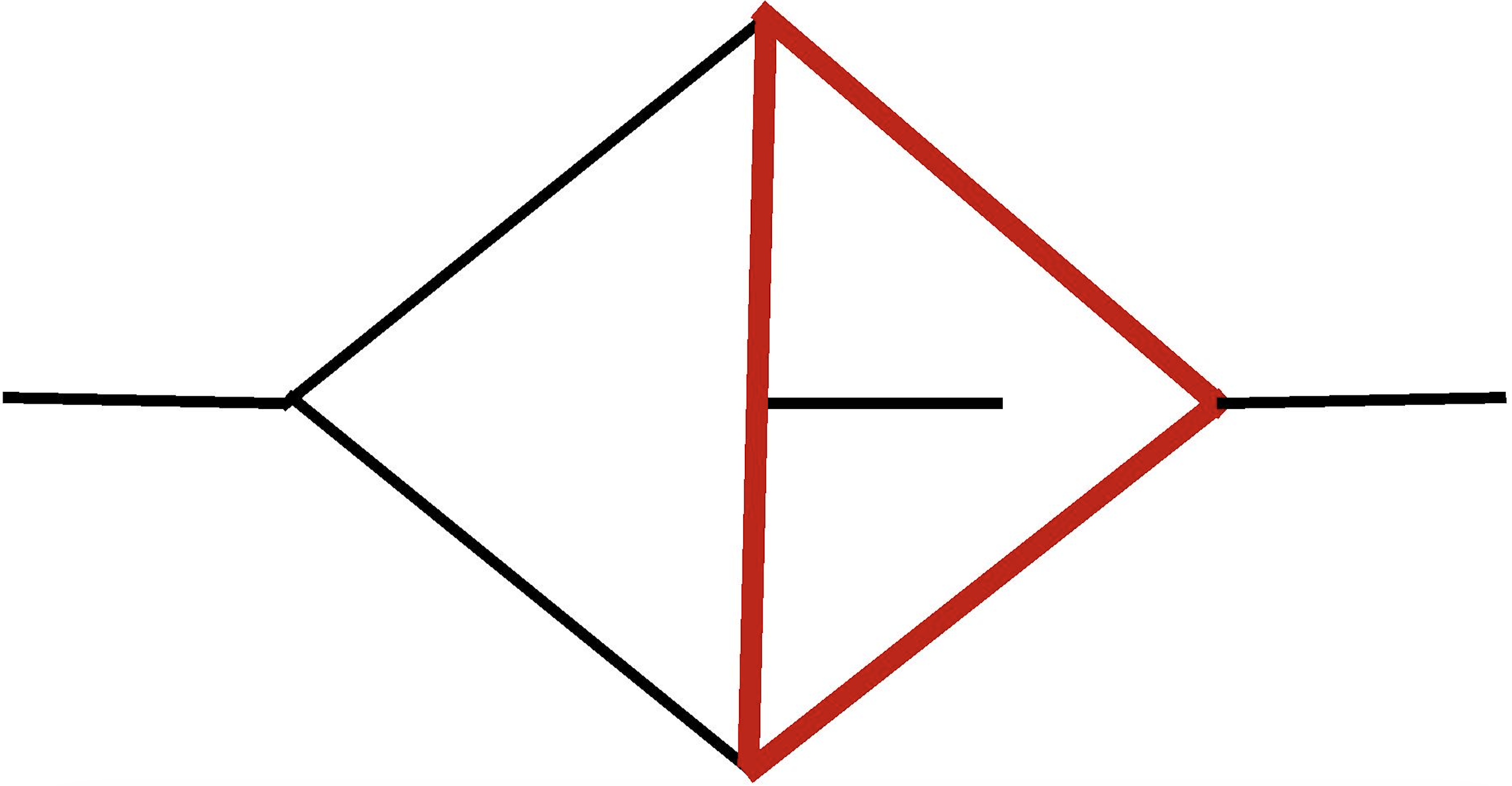}}}
\hspace{0.8cm}
\subfloat[(b)]{\adjustbox{valign=c}{\includegraphics[width = 2.1in, height=1.2in]{Topo12.png}}}
\caption{The non-planar sub-topology and the top-topology where the red and black lines respectively denote massive and massless propagators.}
\label{fig:familyE}
\end{center}
\end{figure}

The maximal cut for the top-sector integral $ I_{111111100}$, as shown in figure.~\ref{fig:familyE}(b),  is given by
\begin{equation}
\label{eq:maxtop}
   \frac{16}{\pi^4} \int_\mathcal{C} \frac{d P}{s\; (s + t + P)\; \sqrt{P}\; \sqrt{
 s + P}\;  \sqrt{-4 m_t^2 s + sP  + P^2} } + O(\epsilon)
\end{equation}
 in $d$-dimensions, where the leading term in $\epsilon$ corresponds to the 4-dimensional part. The integration is over the contour $\mathcal{C}$ in the remaining Baikov variable $P$, after incorporating all the delta distributions that arise from putting all the propagators on-shell. The contour can be chosen to lie between any pair of roots of the polynomial in the denominator. 
On the other hand, the maximal cut of the elliptic sub-sector integral $I_{101111100}$, as shown in figure.~\ref{fig:familyE}(a), is given by
\begin{equation}
\label{eq:maxsubsec}
    \frac{8}{\pi^3}\int_\mathcal{C}  \frac{dP}{ s\; \sqrt{P}\; \sqrt{s + P}\;
   \sqrt{-4 m_t^2 s + s P + P^2}} + O(\epsilon).
\end{equation}
Looking at the equations~(\ref{eq:maxtop}) and~(\ref{eq:maxsubsec}), it becomes evident that both of them contain the same quartic square root, while the $t$-dependence factors out in equation~(\ref{eq:maxtop}). 
A square root $\sqrt{Q(x)}$, where $Q(x)$ is a quartic polynomial with distinct zeroes, defines an elliptic curve. The occurrence of the same quartic square roots strongly suggests the appearance of a single elliptic curve in sectors $J_{31}$-$J_{36}$, instead of two different elliptic curves, a fact that has been overlooked in earlier studies~\cite{Becchetti:2023wev,Xu:2018eos}. This information is useful for the construction of an $\epsilon$-factorised differential equation for the top-sector. Particularly, we find that the $t$ dependence completely decouples after using the factorization properties of the corresponding Picard-Fuchs operator~\cite{Muller-Stach:2012tgj,Adams:2017tga} from the sector that is still coupled at the order $\epsilon^0$. We discuss it in detail in the next sections.

\subsection{Transformation matrix to obtain an $\epsilon$-form for the top-sector on the maximal cut}
\label{subsec:transformation_top}
In this section, we first discuss the steps to obtain the special linear form~\eqref{linearDE} for the differential equations on the maximal cut corresponding to the elliptic top sector integrals $(J_{33},..,J_{36})$ defined in equation~\eqref{eq:startingDE}. Later on, we comment on obtaining the $\epsilon$-factorized canonical form~\eqref{epsDE} from the special linear form~\eqref{linearDE} for these integrals. Following \cite{Adams:2017tga}, we first act on the differential equation for this $4 \times 4$ block with a transformation matrix that decouples the elliptic curve at the order $\epsilon^0$. As a side product, this procedure already transforms the first 2$\times$2 block for the partial differential equation with respect to $t$ in a $\epsilon$-factorized form. In particular, we use the transformation matrix

\begin{equation}
U_0=\small{
\left(\begin{array}{cccc}
0 & -\frac{2 s}{(16+s)(s+2 t)} & \frac{16 t+2 s(4+s+t)}{(16+s)(s+2 t)} & \frac{2 s}{(16+s)(s+2 t)} \\
0 & \frac{8 s(12+s)}{(16+s)^2(s+2 t)} & \frac{-4 s(112+s(33+2 s))-8(112+s(21+s)) t}{(16+s)^2(s+2 t)} & -\frac{8 s(12+s)}{(16+s)^2(s+2 t)} \\
s \sqrt{t} \sqrt{s+t} \sqrt{s(-4+t)+t^2} & 0 & 0 & 0 \\
0 & 0 & 0 & s
\end{array}\right)}
\end{equation}
to get
\begin{equation}
\Tilde{A}^{(0)}_{s,U_0}=
\begin{pmatrix}
\frac{2}{s} & \frac{1}{s} & 0 & 0\\
-\frac{4 (49 + 5 s)}{s (16 + s)} & \frac{-80 - 7 s}{s (16 + s)} & 0 & 0\\
  \frac{\sqrt{t}(-112 s - 8 s^2 + 27 s t + 2 s^2 t + 26 t^2 + 2 s t^2)}{
  \sqrt{s + t} \sqrt{-4 s + s t + t^2}} & \frac{\sqrt{t} (-64 s - 4 s^2 + 16 s t + s^2 t + 16 t^2 + s t^2)}{2 \sqrt{s + t} \sqrt{-4 s + s t + t^2}}
   & 0 & 0\\
-\frac{s}{2} & 0 & 0 & 0
\end{pmatrix},
\end{equation}
\begin{equation}
\Tilde{A}^{(0)}_{t,U_0}=
\begin{pmatrix}
0 & 0 & 0 & 0\\
0 & 0 & 0 & 0\\
  -\frac{s \left(2 s^2 t-9 s^2+2 s t^2+24 s t-112 s+24 t^2\right)}{\sqrt{t} \sqrt{s+t} \sqrt{s t-4 s+t^2}} & -\frac{s (s+16) \sqrt{s t-4 s+t^2}}{2 \sqrt{t} \sqrt{s+t}}
   & 0 & 0\\
0 & 0 & 0 & 0
\end{pmatrix},
\end{equation}
where $\Tilde{A}^{(0)}_{s,U_0}$ and $\Tilde{A}^{(0)}_{t,U_0}$ are the transformed differential equation matrices with respect to $s$ and $t$, respectively, obtained after transforming with $U_0$. The elliptic nature of the Feynman integrals becomes evident when we look at the upper left $2\times2$ block in $\Tilde{A}^{(0)}_{s,U_0}$, which is still coupled at $\epsilon^0$. Indeed, if we compute the Picard-Fuchs operator at the order  $\epsilon^0$ denoted by $L_0$ for the first element from the transformed basis, i.e. the combination of master integrals that we obtain after acting with the transformation matrix $U_0$, which for our case is
\begin{align}
I_1 \;&=\frac{2 \epsilon^4}{(16 + s) (s + 
   2 t)}\;\big(s I_{1, 1, 1, 1, 1, 1, 1, -1, -1}\; - 
   s I_{1, 1, 1, 1, 1, 1, 1, 0, -2}\; + \nonumber \\
   &\hspace{4cm}(4 s + 
   s^2+ 
   8 t+ 
   s t)I_{1, 1, 1, 1, 1, 1, 1, 0, -1} \big)  , 
\end{align} we obtain
\begin{equation}
   L_0 = \partial^2_s - \left( -\frac{4}{s} -\frac{2}{16+s}\right) \partial_s +\frac{6 (6 + s)}{s^2 (16 + s)}.
\end{equation}
This second order differential operator is associated to an elliptic curve and its two solutions can be chosen as
\begin{align}
    &\psi_0(s)= \frac{32\; E(-\frac{s}{16})}{\pi  s^{3/2} (s+16)},
    \label{period1}\\
    &\psi_1(s)= \frac{32\; \left(E(\frac{s}{16}+1)-K(\frac{s}{16}+1)\right)}{s^{3/2} (s+16)}
    \label{period2},
\end{align}
where $\psi_0(s)$ is a holomorphic function multiplied by $\frac{1}{s^{3/2}}$ and $\psi_1(s)$ is the first logarithmic solution. 
Now we extend the method of~\cite{Pogel:2022vat} with minor modifications to multiple scales, in order to factorize $\epsilon$ from this $2\times2$ block.  In particular, we look for new master integrals defined in the following way
\begin{align}
    &M_1=\frac{1}{h(s,t)}I_1, \nonumber \\
    &M_2=\frac{g_1(s,t)}{\epsilon}\frac{d}{ds}M_1+\frac{g_2(s,t)}{\epsilon}\frac{d}{dt}M_1-f_1(s,t)M_1.
    \label{unknowbasis}
\end{align}
The above is the most general ansatz with four rational functions:  $h(s,t), \; g_1 (s,t),\; g_2 (s,t),$ ${\text{and}}\; f_1(s,t) $, that one can write thinking about the Hodge filtration of a Hodge structure~\cite{Bonisch:2021yfw}. With this ansatz, the differential equation matrix $\Tilde{A}$ for the first 2 $\times$ 2 block transforms again as in equation~(\ref{trafoDE}). We find the unknown functions in (\ref{unknowbasis}) by requiring that the transformed matrix $\Tilde{A}$ has terms proportional to $\epsilon$ and as simplest as possible $\epsilon^0$ terms. This is done by choosing
\begin{align}
    &h(s,t)= \frac{1}{s^2+16 s},\\
    &g_1(s, t)=0,\\
    &g_2(s, t)=\frac{16 \;s \;(s + 2 t)^2}{16 \;t + s\; (8 + t)},\\
    &f_1(s, t)=\frac{16\; s^3\;(s + 2 t)}{t\; (s + t)\; (16\; t + s\; (8 + t))}.
\end{align}
So the next transformation matrix attains the form
\begin{equation}
U_1=
    \begin{pmatrix}
        s\; (16 + s) & 0\\
        -64\; s^2 \;(12 + s) & -16\; s^2\; (16 + s)
    \end{pmatrix}.
\end{equation}
Acting with another $2\times2$ rotation, where the entries of the matrix are constrained to eliminate the $\epsilon^0$ terms, we $\epsilon$-factorize the $2\times2$ block. This transformation is given by
\begin{equation}
U_2=
    \begin{pmatrix}
        \sqrt{s}\; K(\frac{s}{16}+1) & \frac{1}{8 \sqrt{s}}\bigg(E(\frac{s}{16}+1)-K(\frac{s}{16}+1)\bigg)\\
        \frac{2 \sqrt{s}}{\pi}\; K(-\frac{s}{16}) & -\frac{1}{4 \pi \sqrt{s}}{E(-\frac{s}{16})}
        \end{pmatrix}.
\end{equation}
Here the appearance of elliptic functions is not surprising given that the associated geometry is elliptic in nature. These entries can indeed be rewritten in terms of the two periods $\psi_0$ and $\psi_1$ defined in equations~(\ref{period1}) and (\ref{period2}) and their derivatives in the following way
\begin{equation}
U_2=\frac{s^3 (s+16)^2}{16} 
    \begin{pmatrix}
         s \psi'_1 -2 \psi_1 & - \psi_1\\
        -2  (s \psi'_0-2 \psi_0) & 2  \psi_0
    \end{pmatrix}.
\end{equation}
After fixing the $2 \times 2$ sub-block, we proceed to $\epsilon$-factorize the remaining $4 \times 4$ differential matrix. It is important to note that at this stage, we already obtain the special linear form for our differential equations, which enables us to conveniently write the analytic results for the elliptic top-sector (on the cut) in terms of iterated integrals. However, as mentioned before, the results at a particular $\epsilon$-order do not have the same length of iterated integrals. Let us now look at how to obtain an $\epsilon$-form for the full $4 \times 4$ matrix. After applying the rotations $U_1$ and $U_2$, $\Tilde{A}$ transforms accordingly as in equation~(\ref{trafoDE}). We find it useful to examine the new differential equation matrix, $\Tilde{A}_{U_2U_1}$, based on its scale dependence, i.e. we analyze $\Tilde{A}_{U_2U_1,s}$ and $\Tilde{A}_{U_2U_1,t}$ individually. We find that $\Tilde{A}_{U_2U_1,s}$ has $\epsilon^0$ terms only at the positions (3,1), (3,2), (4,1) and (4,2) whereas $\Tilde{A}_{U_2U_1,t}$ has non-zero components only at positions (3,1) and (3,2). With this information,  we construct another transformation matrix as follows:
\begin{equation}
U_3=
    \begin{pmatrix}
        1 & 0 & 0 & 0\\
        0 & 1 & 0 & 0\\
        0 & 0 & 1 & 0\\
        -\frac{1}{2} s^2 (s+12) & -\frac{1}{8} s^2 (s+16) & 0 & 1        
    \end{pmatrix},
\end{equation}
where the components are fixed by requiring that the $\epsilon^0$ terms at positions (4,1) and (4,2) cancel each other out. The last transformation $U_4$, to fully $\epsilon$-factorize the $4 \times 4$ block of the differential equation system for the top-sector, is obtained in a similar way by constructing another non-trivial transformation for the elements at position (3,1) and (3,2) through
%%I DON'T KNOW IF I TRIED WITH VARYING EPS
\begin{equation}
U_4=
    \begin{pmatrix}
        1 & 0 & 0 & 0\\
        0 & 1 & 0 & 0\\
        f(s,t) & g(s,t) & 1 & 0\\
        0 & 0 & 0 & 1 
    \end{pmatrix}.
\end{equation}
The definitions of $f(s,t)$ and $g(s,t)$ are given in the appendix~\ref{sec:appendeixfg}. Furthermore, in appendix~\ref{sec:appendixtorus}, we highlight the significance of studying the marked point on the torus to simplify these functions. This is obtained after establishing a correspondence between the elliptic curve and a torus. 
 
In conclusion, we obtain both a completely $\epsilon$-factorized as well as a special linear form~(\ref{linearDE}) for the $4 \times 4$  differential equation block corresponding to the elliptic top sector. We also find that the analytic one-forms are simpler if we only transform the starting block for the top-sector (the one obtained with $J_i$ in equation~(\ref{eq:startingDE})), from $U_0$ till $U_3$, rather than proceeding to the final transformation matrix $U_4$.
Hence this is the choice made in our supplied transformation files. We also write the corresponding differential equation matrix $\Tilde{A}_{f(i,j)}$, with $f\in\{s,t\}$, for the elliptic top sector explicitly in appendix~\ref{sec:appendixoneform}. The corresponding differential one-forms are expressed in terms of the periods and their derivatives given in equations~(\ref{period1}) and~(\ref{period2}). All the elements are proportional to $\epsilon$, apart from $\Tilde{A}_{f(3,1)}$ and $\Tilde{A}_{f(3,2)}$.
 
\subsubsection*{Analytic results through Chen's iterated integrals}
\label{sec:Analyt-Num-results}
The primary focus of this article revolves around addressing the algebraic complexity arising from two elliptic sectors within a multi-variate non-planar Feynman integral family. As evident from the previous section, we made a comprehensive investigation of all the elliptic sectors to manage this complexity. However, as a secondary outcome, we also provide a choice of basis for the full system of differential equations as ancillary materials with this article. This basis can be utilised to express all the appearing Feynman integrals analytically through Chen's iterated integrals, albeit with non-uniform length, at each order in $\epsilon$. This choice has been obtained as follows.  Typically, following the bottom-top approach, selecting an appropriate basis for the top sector on the maximal cut, with all the sub-sector integrals already in a canonical form, also simplifies the off-diagonal differential equation contributions to the top-sector. This is true especially concerning the $\epsilon$ expansion behaviour. However, we observe that this is not the case here. In particular, we additionally fix the $\epsilon$-behaviour of these off-diagonal elements, to limit the one-forms to have only finite expansions in $\epsilon$. We provide a choice of basis that is arranged in a lower block-triangular form which nevertheless enables one to express the analytic results though iterated integrals, after solving the differential equations order by order in $\epsilon$. We would also like to mention that after obtaining the system of differential equations with the provided choice of basis for the full system, further transformations~ \cite{Meyer:2017joq} can be carried out, to fix only the sub-sector contributions to the top-sector, to obtain a completely $\epsilon$-factorised differential equations for the full $36 \times 36$ system. However, in this case, the more transformations carried out, the more intricate the structure of the kernels within the iterated integrals becomes.

\section{Conclusions and outlook}
\label{sec:conclusions}
In this paper, we studied the analytic structure of non-planar Feynman integrals crucial for computing NNLO QCD corrections in diphoton and dijet production. We presented an extensive discussion on deriving an $\epsilon$-factorized form of the differential equations for the elliptic top sector, leveraging the geometric properties of the Feynman integral and insights obtained from the maximal cuts. Contrary to the earlier studies and despite the presence of an elliptic sub-topology, we identify only one elliptic curve in the entire system instead of two distinct elliptic curves. With this information, we exploit the properties of the Picard-Fuchs operator to find an $\epsilon$-factorised differential equation of the top-sector on the maximal cut. Additionally, we adapted the algorithm outlined in~\cite{Pogel:2022vat} for our multi-variate elliptic system, enabling the derivation of appropriate ansätze for the top sector. At the end, we also briefly discussed how to obtain analytic results for the elliptic top sector integrals in terms of Chen's iterated integrals using our choice of basis.

Looking ahead, although significant progress has been achieved in understanding the analytic structure of this family of Feynman integrals, further investigation is clearly warranted. Specifically, it would be intriguing to obtain a simpler definition of a good function basis for these integrals, incorporating information from both the maximal and sub-maximal cuts from the beginning. This approach would result in requiring fewer transformations to obtain the $\epsilon$-factorized differential equation for the entire system, thereby yielding less complicated kernels in the analytic results. Furthermore, the numerical evaluation of these elliptic kernels needs to be explored.

\section*{Acknowledgements}
We would like to thank Claude Duhr and Vasily Sotnikov for comments on the manuscript. EC would also like to thank Christoph Dlapa for interesting discussions at the workshop `MathemAmplitudes 2023: QFT at the Computational Frontier'. SM would like to thank Yu Jiao Zhu for insightful discussions. The work of EC and SM is funded by the ERC grant 101043686 ‘LoCoMotive’.  The work of TA is supported by Deutsche Forschungsgemeinschaft (DFG) through the Research Unit FOR
2926, Next Generation perturbative QCD for Hadron Structure: Preparing for the Electron-ion collider,
project number 409651613.  MK would like to acknowledge
financial support from IISER Mohali for this work.
\newpage
\clearpage
\section{Appendix}

\subsection{The differential one-forms for the elliptic top sector}
\label{sec:appendixoneform}
The matrix valued differential one-forms for the top-sector mentioned in section~\ref{subsec:transformation_top} has the following form:
{\small{    
\begin{align}
\label{eq:matrixoneforms}
    \Tilde{A}_{f(1,1)}=&\frac{\epsilon}{512}   (2 \text{$\psi_0$} \left(\text{$\psi_1$}' (\text{ds}\; w_{22}-\text{dt}\; w_{3})-2 \text{$\psi_1$} (\text{ds}\; w_{21}+\text{dt}\; w_{1})\right)+\nonumber\\
    &\text{$\psi_0$}' \left(\text{$\psi _1$}' (\text{ds}\; w_{24}-\text{dt}\; w_{2})+2 \text{$\psi_1$} (\text{ds}\; w_{23}-\text{dt}\; w_{4})\right)+512 \text{dt}\; w_{16})\,,\nonumber\\
    \Tilde{A}_{f(1,2)}=&\frac{\epsilon}{1024}  \left(4 \text{ds}\; w_{25}-\text{dt}\; w_{5} \text{$\psi_1$} \text{$\psi_1$}'\right)\,,\nonumber\\
    \Tilde{A}_{f(1,3)}=&\frac{\epsilon}{8}   \left(\text{$\psi_1$}' (\text{ds}\; w_{31}-\text{dt}\; w_{11})+2 \text{$\psi_1$} (\text{ds}\; w_{26}-\text{dt}\; w_{6})\right)\,,\nonumber\\
    \Tilde{A}_{f(1,4)}=&-\frac{\epsilon }{4}  \left(\text{$\psi_1$}' (\text{dt}\; w_{12}-\text{ds}\; w_{32})+2 \text{$\psi_1$} (\text{ds}\; w_{27}+\text{dt}\; w_{7})\right)\,,\nonumber\\
    \Tilde{A}_{f(2,1)}=&\frac{\epsilon }{256}  \left(\text{dt}\; w_{5} \text{$\psi_0$} \text{$\psi_0$}'-4 \text{ds}\; w_{25}\right)\,,\nonumber\\
    \Tilde{A}_{f(2,2)}=&\frac{\epsilon}{512}   (2 \text{$\psi_0$} \left(2 \text{$\psi_1$} (\text{ds}\; w_{21}+\text{dt}\; w_{1})+\text{$\psi_1$}' (\text{dt}\; w_{4}-\text{ds}\; w_{23})\right)\nonumber\\
    &+\text{$\psi_0$}' \left(\text{$\psi_1$}' (\text{dt}\; w_{2}-\text{ds}\; w_{24})+\text{$\psi _1$} (2 \text{dt}\; w_{3}-2 \text{ds}\; w_{22})\right)+512 \text{dt}\; w_{16})\,,\nonumber\\
    \Tilde{A}_{f(2,3)}=&-\frac{\epsilon}{4}   \left(\text{$\psi_0$}' (\text{ds}\; w_{31}-\text{dt}\; w_{11})+2 \text{$\psi_0$} (\text{ds}\; w_{26}-\text{dt}\; w_{6})\right)\,,\nonumber\\
    \Tilde{A}_{f(2,4)}=&\frac{\epsilon}{2}   \text{$\psi_0$}' (\text{dt}\; w_{12}-\text{ds}\; w_{32})+\epsilon  \text{$\psi_0$} (\text{ds}\; w_{27}+\text{dt}\; w_{7})\,,\nonumber\\
    \Tilde{A}_{f(3,1)}=&\frac{1}{64} \left(2 \text{$\psi_0$} (\text{ds}\; (2 w_{29}+w_{30} \epsilon )-\text{dt}\; (w_{10} \epsilon +w_{9}))+\text{$\psi_0$}' (2 \text{ds}\; w_{34}+\text{ds}\; w_{35} \epsilon -2 \text{dt}\; w_{14}-\text{dt}\; w_{15} \epsilon )\right)\,,\nonumber\\
    \Tilde{A}_{f(3,2)}=&\frac{1}{128} \left(2 \text{$\psi_1$} (\text{ds}\; (2 w_{29}+w_{30} \epsilon )-\text{dt}\; (w_{10} \epsilon +w_{9}))+\text{$\psi_1$}' (2 \text{ds}\; w_{34}+\text{ds}\; w_{35} \epsilon -2 \text{dt}\; w_{14}-\text{dt}\; w_{15} \epsilon )\right)\,,\nonumber\\
    \Tilde{A}_{f(3,3)}=&\epsilon  (4 \text{ds}\; w_{36}-4 \text{dt}\; w_{17})\,,\nonumber\\
    \Tilde{A}_{f(3,4)}=&\epsilon  (2 \text{ds}\; w_{37}-2 \text{dt}\; w_{18})\,,\nonumber\\
    \Tilde{A}_{f(4,1)}=&\frac{ \epsilon }{128} \left(\text{$\psi_0$}' (\text{ds}\; w_{33}+4 \text{dt}\; w_{13})+2 \text{$\psi_0$} (\text{ds}\; w_{28}+4 \text{dt}\; w_{8})\right)\,,\nonumber\\
    \Tilde{A}_{f(4,2)}=&\frac{\epsilon}{256}   \left(\text{$\psi_1$}' (\text{ds}\; w_{33}+4 \text{dt}\; w_{13})+2 \text{$\psi_1$} (\text{ds}\; w_{28}+4 \text{dt}\; w_{8})\right)\,,\nonumber\\
    \Tilde{A}_{f(4,3)}=&\frac{ \epsilon }{2} (\text{dt}\; w_{19}-\text{ds}\; w_{38})\,,\nonumber\\
    \Tilde{A}_{f(4,4)}=&\epsilon  (\text{dt}\; w_{20}-\text{ds}\; w_{39})\,,\nonumber\\
\end{align}
}}
where
{\small{ 
\begin{align}
    & w_{1}=\frac{s^3 (s+8) (s+12) (s+16)}{s+2 t}\,, \qquad
     w_{2}=\frac{s^5 (s+16)^3}{s+2 t}\,, \qquad 
     w_{3}=\frac{s^4 (s+8) (s+16)^2}{s+2 t}\,,\nonumber\\
    & w_{4}=\frac{s^4 (s+12) (s+16)^2}{s+2 t}\,,\qquad
     w_{5}=\frac{s^3 (s+16) (2 s (s+8) (s+16)+2 s (s+12) (s+16))}{s+2 t}\,,\nonumber\\
    & w_{6}=\frac{s^2 (s+16) \left(s^2+(s+12) t^2+(s+12) s t\right)}{\sqrt{t} \sqrt{s+t} (s+2 t)^2 \sqrt{s (t-4)+t^2}}\,,\qquad
     w_{7}=\frac{s^3 (s+8)}{(s+2 t)^2}\,, \qquad
     w_{8}=s (s+8)\,,\nonumber \\
    & w_{9}=\frac{s \left(2 s^2 (t-5)+2 s \left(t^2+8 t-48\right)+16 t^2\right)}{\sqrt{t} \sqrt{s+t} \sqrt{s (t-4)+t^2}}\,,
     w_{10}=\frac{s \left(3 s^3+2 s^2 (6 t+22)+12 s (t+4)^2+96 t^2\right)}{\sqrt{t} \sqrt{s+t} \sqrt{s (t-4)+t^2}},\nonumber\\
    & w_{11}=\frac{s^3 (s+16)^2 \sqrt{t} \sqrt{s+t}}{(s+2 t)^2 \sqrt{s (t-4)+t^2}}\,,\qquad
     w_{12}=\frac{s^4 (s+16)}{(s+2 t)^2}\,,\qquad
     w_{13}=s^2 (s+16)\,,\nonumber\\
    & w_{14}=\frac{s^2 (s+16) \sqrt{s t-4 s+t^2}}{\sqrt{t} \sqrt{s+t}}\,,\qquad
     w_{15}=\frac{s^2 (s+16) \left(3 s^2+12 s t+16 s+12 t^2\right)}{\sqrt{t} \sqrt{s+t} \sqrt{s (t-4)+t^2}}\,,\nonumber\\
    & w_{16}=\frac{s^2}{t (s+t) (s+2 t)}\,,\qquad
     w_{17}=\frac{s^3 (t-1)+4 s^2 t^2+6 s t^3+3 t^4}{t (s+t) (s+2 t) \left(s (t-4)+t^2\right)}\,,\nonumber \\
    & w_{18}=\frac{3 s^2+8 s t+8 t^2}{\sqrt{t} \sqrt{s+t} (s+2 t) \sqrt{s (t-4)+t^2}}\,,\qquad
     w_{19}=\frac{s+2 t}{\sqrt{t} \sqrt{s+t} \sqrt{s (t-4)+t^2}}\,,\quad
     w_{20}=\frac{s+2 t}{t (s+t)}\,,\nonumber \\
    & w_{21}=\frac{s^2 ((384-s (s (s+16)+24)) t+2 s (s+10) (5 s+48))}{s+2 t}\,,\nonumber\\
    & w_{22}=\frac{s^3 (s+16) \left(s^3 (t-11)+s^2 (t-14) (t+8)+4 s (t-40) t-64 t^2\right)}{(s+t) (s+2 t)}\,,\nonumber\\
    & w_{23}=\frac{s^3 (s+16) \left(-s^2 (7 s+64)+(s+4) (s+8) t^2+(s (s+4)-48) s t\right)}{(s+t) (s+2 t)}\,,
     w_{24}=\frac{s^5 (s+16)^2 (t-8)}{s+2 t},\nonumber\\
    & w_{25}=\frac{s^3 (s+16) (s ((s+8) t-9 s-88)-16 t)}{s+2 t}\,,\nonumber \\
    & w_{26}=\frac{s \sqrt{t} \left(-4 (s+8) s^2+(s (s+18)+96) t^2+(s (s+13)+48) s t\right)}{\sqrt{s+t} (s+2 t)^2 \sqrt{s (t-4)+t^2}}\,,\qquad 
     w_{27}=\frac{s^2 (s-(s+6) t)}{(s+2 t)^2}\,,\nonumber\\
    & w_{28}=(s+8) (s-2 t)\,,\qquad 
     w_{29}=\frac{\sqrt{t} \left((s+10) t^2+s (s+11) t-4 s (s+12)\right)}{\sqrt{s+t} \sqrt{s (t-4)+t^2}}\,,\nonumber\\
    & w_{30}=\frac{\sqrt{t} \left(2 (s+12) (3 s+16) t^2+3 s (s (s+4)-64) t+64 s (5 s+48)\right)}{(s+16) \sqrt{s+t} \sqrt{s (t-4)+t^2}}\,,\nonumber\\
    & w_{31}=\frac{s^2 (s+16) \sqrt{t} \sqrt{s+t} (s (t-4)+8 t)}{(s+2 t)^2 \sqrt{s (t-4)+t^2}}\,,\qquad 
     w_{32}=\frac{s^3 (s+16) t}{(s+2 t)^2}\,,\qquad
     w_{33}=s (s+16) (s-2 t)\,,\nonumber\\
    & w_{34}=\frac{s (s+16) \sqrt{t} \sqrt{s (t-4)+t^2}}{\sqrt{s+t}}\,,\qquad
     w_{35}=\frac{s \sqrt{t} \left(3 s^2 t+2 (3 s+32) t^2+256 s\right)}{\sqrt{s+t} \sqrt{s (t-4)+t^2}}\,,\nonumber\\
    & w_{36}=\frac{t \left(3 s^2+s (t+2) t+t^3\right)}{s (s+t) (s+2 t) \left(s (t-4)+t^2\right)}\,,\qquad 
     w_{37}=\frac{\sqrt{t} \left(-s^2 (t-16)+2 s t (t+16)+64 t^2\right)}{s (s+16) \sqrt{s+t} (s+2 t) \sqrt{s (t-4)+t^2}}\,,\nonumber\\
    & w_{38}=\frac{t^{3/2}}{s \sqrt{s+t} \sqrt{s (t-4)+t^2}}\,,\qquad 
    w_{39}=\frac{s+2 t}{s (s+t)}\,.\nonumber 
\end{align}
}}    
In equation~(\ref{eq:matrixoneforms}), we have suppressed the explicit dependence of the periods and their derivatives on $s$.

\subsection{The definitions of the entries in $U_4$}
\label{sec:appendeixfg}
In this section, we first provide the definitions of $f(s,t)$ and $g(s,t)$ mentioned in section~\ref{subsec:transformation_top}, that appears in the last transformation matrix $U_4$ needed to transform the differential equations corresponding to the top-sector to an $\epsilon$-factorised form. Later we also perform a study of the maximal cut again to suggest a change of variables that simplifies these functions. 
{\small{
\begin{align}
    f(s,t)&= \frac{s}{\sqrt{t}\sqrt{s+t}} \Bigg\{  
    2 (s+12) (s+t) \sqrt{s (t-4)+t^2} \nonumber\\
    &-\frac{2 \sqrt{2} \sqrt{s} \sqrt{t}(5 s+64)}{\sqrt{\frac{\big(s+\sqrt{s} \sqrt{s+16}+8\big)}{s+t}}}  \;\;F\bigg(\sin ^{-1}\bigg(\frac{\sqrt{1+\frac{(t-8) \sqrt{s}}{t\sqrt{s+16}}}}{\sqrt{2}}\bigg)|\frac{2 \sqrt{s} \sqrt{s+16}}{s+\sqrt{s} \sqrt{s+16}+8}\bigg)\nonumber\\
    &-
    \frac{\sqrt{2} (s+12) \sqrt{s} (s+t)}{\sqrt{\frac{s+t}{\left(s+\sqrt{s} \sqrt{s+16}+8\right) t}}}  \;\;E\bigg(\sin ^{-1}\bigg(\frac{\sqrt{1+\frac{(t-8) \sqrt{s}}{t\sqrt{s+16}}}}{\sqrt{2}}\bigg)|\frac{2 \sqrt{s} \sqrt{s+16}}{s+\sqrt{s} \sqrt{s+16}+8}\bigg) \Bigg\},
\end{align}
}}
{\small{
\begin{align}
    g(s,t)&=-\frac{s (s+16)}{4 \sqrt{t}\sqrt{s+t}} \Bigg\{ \big(-2 (s+t) \sqrt{{s (t-4)+t^2}}\nonumber\\
    &    
    +\frac{8 \sqrt{2} \sqrt{s} \sqrt{t}}{\sqrt{\frac{\big(s+\sqrt{s} \sqrt{s+16}+8\big)}{s+t}}}  \;\;F\bigg(\sin ^{-1}\bigg(\frac{\sqrt{1+\frac{(t-8) \sqrt{s}}{t\sqrt{s+16}}}}{\sqrt{2}}\bigg)|\frac{2 \sqrt{s} \sqrt{s+16}}{s+\sqrt{s} \sqrt{s+16}+8}\bigg)\nonumber\\
    &+\frac{\sqrt{2} \sqrt{s} (s+t)}{{\sqrt{\frac{s+t}{\big(s+\sqrt{s} \sqrt{s+16}+8\big) t}}}} \;\;E\bigg(\sin ^{-1}\bigg(\frac{\sqrt{1+\frac{(t-8) \sqrt{s}}{t\sqrt{s+16}}}}{\sqrt{2}}\bigg)|\frac{2 \sqrt{s} \sqrt{s+16}}{s+\sqrt{s} \sqrt{s+16}+8}\bigg)\Bigg\},
\end{align}
}}
where $F(\cdot|\cdot)$ and $E(\cdot|\cdot)$ are respectively the elliptic integral of the first and second kind:
\begin{equation}
    F(\phi|k^2)=\int_0^{\sin{\phi}}\frac{dt}{\sqrt{(1-t^2)(1-k^2t^2)}}, \qquad E(\phi|k^2)=\int_0^{\sin{\phi}} \sqrt{\frac{1-k^2t^2}{1-t^2}} dt\,.
\end{equation}
\subsection{The study of the marked point on the torus}
\label{sec:appendixtorus}
Below we study the correspondence between the elliptic curve and a torus to find a possible change of variables that enables us to express  the functions $f(s,t)$ and $g(s,t)$ in a more compact notation. This might be useful when aiming to express the results through eMPLs in future works.  From the maximal cut in equation~(\ref{eq:maxtop}) we can see that there is a pole in $P$ at $s+t$. A pole in the maximal cut corresponds to a marked point in the torus. The map between an elliptic curve and the corresponding torus is given by Abel's map and as done in~\cite{Giroux:2022wav, Delto:2023kqv} we can compute how the pole is marked on the torus
\begin{equation}
    Z_{x_p}=\frac{F(\sin^{-1}{u_{x_p}}|k)}{2 K(k)},
\end{equation}
where
\begin{equation}
    k^2=\frac{r_{23}\;r_{14}}{r_{13}r_{24}}, \qquad r_{ij}=r_i-r_j
\end{equation}
and
\begin{equation}
    u_{x_p}=\sqrt{\frac{(x_p-r_1)(r_2-r_4)}{(x_p-r_2)(r_1-r_4)}}\,.
\end{equation}
In the above equation, $r_i$ are the roots of the elliptic curve and $x_p$ is the pole on the maximal cut. 

With the roots of our elliptic curve
\begin{equation}
    r_1=\frac{1}{2} \left(-s-\sqrt{s} \sqrt{s+16}\right), \quad r_2=-s, \quad r_3=0, \quad r_4=\frac{1}{2} \left(\sqrt{s} \sqrt{s+16}-s\right), \quad x_p=-s-t,
\end{equation}
we get the following $k^2$ and $u_{x_p}$
\begin{equation}
    k^2=\frac{2 \sqrt{s} \sqrt{s+16}}{s+\sqrt{s} \sqrt{s+16}+8}, \qquad u_{x_p}=\frac{\sqrt{1+\frac{(t-8) \sqrt{s}}{t\sqrt{s+16}}}}{\sqrt{2}},
\end{equation}
and the following marked point on the torus
\begin{equation}
    Z_{x_p}=\frac{F\bigg(\sin ^{-1}\bigg(\frac{\sqrt{1+\frac{(t-8) \sqrt{s}}{t\sqrt{s+16}}}}{\sqrt{2}}\bigg)|\frac{2 \sqrt{s} \sqrt{s+16}}{s+\sqrt{s} \sqrt{s+16}+8}\bigg)}{2 K\bigg(\frac{2 \sqrt{s} \sqrt{s+16}}{s+\sqrt{s} \sqrt{s+16}+8}\bigg)}.
\end{equation}
We notice that this is the same integral that appears in the expression for $f(s,t)$ and $g(s,t)$, so we rewrite them using these variables and notice that their expression simplifies
{\small{
\begin{align}
    &f(s,t)= \frac{2 s (s+12) \sqrt{s+t} \sqrt{s (t-4)+t^2}}{\sqrt{t}}\nonumber \\
    &+\frac{\sqrt{2} s^{3/2}}{\sqrt{s+\sqrt{s} \sqrt{s+16}+8}}\left(4 (5 s+64) K(k) Z_{x_p}+(s+12) \left(s+\sqrt{s} \sqrt{s+16}+8\right) E\left(\left.\sin ^{-1}(u_{x_p})\right|k\right)\right),\\
    &g(s,t)=\frac{(s+16) s \sqrt{s+t} \sqrt{s (t-4)+t^2}}{2 \sqrt{t}}\nonumber \\
    &+\frac{\sqrt{2} s^{3/2} (s+16)}{\sqrt{s+\sqrt{s} \sqrt{s+16}+8}}\left(4 K(k) Z_{x_p}+\frac{1}{4} \left(s+\sqrt{s} \sqrt{s+16}+8\right) E\left(\left.\sin ^{-1}(u_{x_p})\right|k\right)\right).
\end{align}
}}

\bibliographystyle{JHEP}
\bibliography{dijet} 
\end{document}